# Permutation-Based Tests of Perfect Ranking

Ehsan Zamanzade[a,*,1], Nasser Reza Arghami[a] and Michael Vock[b]
[a] Department of Statistics, Faculty of Mathematical Sciences, Ferdowsi University of Mashhad, Mashhad, Iran.
[b] Institute of Mathematical Statistics and Actuarial Science, University of Bern, Sidlerstrasse 5, CH-3012 Bern, Switzerland.

## Abstract:

We improve three tests of perfect ranking in ranked set sampling proposed by Li and Balakrishnan (2008) using a permutation approach. This simple way of extending all three concepts to comparisons across different cycles increases the power. Two of the proposed tests are equivalent to tests from the literature, which were derived differently and are therefore generalized by the permutation-based tests.

## 1. Introduction

Ranked set sampling (RSS), which was first introduced by McIntyre (1952) for the estimation of average harvest, is an efficient way for collecting more informative samples from the population under study (and hence making more reliable inferences) provided that the ranking of data (at least in small groups and not necessarily perfect) is much easier and more affordable in comparison with actually measuring them. Ranked set sampling can be done with a balanced or an unbalanced strategy; we concentrate on the balanced case. In balanced ranked set sampling (BRSS), we first draw $k$ random samples of size $k$ and rank them in increasing magnitude (without actually measuring them), then we measure the $i$th smallest observation from the $i$th sample (for $i$ = 1, …, k), and the first sample of size $k$ is obtained. This procedure is repeated $n$ times (cycles) in order to derive a sample of size $nk$ from the BRSS scheme. The resulting measurements are denoted by $X_{[i]ij}$ $i \in \{1, 2, 3, \cdots k\}$, $j \in \{1, 2, 3, \cdots n\}$, where $X_{[i]ij}$ is the actually measured value of the $i$th observation of the $i$th ranked sample (using approximate ranking) in the $j$th cycle. It is worth mentioning that although the $X_{[i]ij}$s are not identically distributed, they are mutually independent and if there were no errors in ranking of samples, they were distributed the same as the $i$th order statistic from a sample of size $k$ from the population. Here, $X_{(i)ij}$, $i \in \{1, 2, 3, \cdots k\}$, $j \in \{1, 2, 3, \cdots n\}$, is the $i$th observation of the $i$th ranked sample in the $j$th cycle when there are no ranking errors (perfect ranking). Unbalanced ranked set sampling differs from BRSS in the way that in each cycle, each sample can have a different size, and that some ranks may be selected more often than others.

---

* Corresponding author.
Email addresses: Ehsanzamanzadeh@yahoo.com (E. Zamanzade), Arghami_nr@yahoo.com (N.R Arghami) and Michael.vock@stat.unibe.ch (M. Vock)
[1] Visiting Researcher at University of Bern, Bern, Switzerland.

Due to the effectiveness and applicability of the RSS technique in many real situations, there has been a lot of research in this area since its introduction. Takahasi and Wakimoto (1968) were the first who formally proved that estimation of the mean based on RSS is more efficient than its estimation based on simple random sampling (SRS). Stokes (1980) proposed an estimate for the population variance in RSS which was improved by Perron and Sinha (2002) in the multi-cycle case and which was improved upon again by Sengupta and Mukhuti (2006) for an exponential population. Estimation of the distribution function in an RSS scheme was proposed by Stokes and Sager (1988). Bohn and Wolfe (1992) and Mahdizadeh and Arghami (2010) used the RSS empirical distribution function in order to introduce Mann–Whitney–Wilcoxon and entropy-based tests in the RSS scheme, respectively. Double ranked set sampling introduced by Al-Saleh and Al-Kadiri (2000) and multistage ranked set sampling proposed by Al-Saleh and Al-Omari (2002) are two extensions of the RSS scheme designed to collect more informative samples in comparison with the RSS scheme. A short but useful description of works based on RSS may be found in Wolfe (2004).

Although statistical inferences based on RSS are more accurate than those based on SRS, this advantage decreases as the ranking error of observations increases. Frey et al (2007) provided two nice examples in which the extent of the ranking error in RSS determines the method of inference. Based on this argument, it seems to be vital to develop some techniques for assessing perfect ranking in data collected by the RSS scheme, but unfortunately, this has not been realized up to recently. Frey et al (2007) and Li and Balakrishnan (2008) independently proposed some methods for testing perfect ranking. Vock and Balakrishnan (2011) improved one of the tests proposed by Li and Balakrishnan (2008) in the multi-cycle case and also compared their proposed test with the tests proposed by Frey et al (2007) and Li and Balakrishnan (2008) by using Monte Carlo simulation under different ranking error scenarios.
The rest of our paper is organized as follows: In Section 2, we introduce our proposed tests based on different permutations of RSS observations. In Section 3, we compare our proposed tests with their leading competitors. Final remarks and recommendations are provided in Section 4.

## 2. Permutation-based test statistics

Suppose that $X_{[1]11}, X_{[2]21}, X_{[3]31}, \cdots, X_{[k]k1}$ are drawn from a one-cycle BRSS scheme. One may expect that if the assumption of perfect ranking is completely satisfied, then the rank of $X_{[i]i1}$ in the sample should be close to the rank of *i*th order statistic in that sample (namely *i*). Therefore it seems reasonable to propose tests of perfect ranking based on the difference between the rank of $X_{[i]i1}$ in the sample and *i*. Thus the following test statistics have been suggested:

$$S_k = \sum_{i=1}^{k} \left(R_i - i\right)^2$$

$$A_k = \sum_{i=1}^{k} \left|R_i - i\right|$$

where $R_i$ is the rank of $X_{[i]i1}$ in the RSS sample.

Furthermore, under the assumption of perfect ranking, $P\left(X_{[j]j1} > X_{[i]i1}\right) > .5,\quad for\ all\ j > i$, thus the following test statistic for perfect ranking can be regarded as the number of pairs of measured observations that violate the expected order:

$$N_k = \sum_{i=1}^{k-1} \sum_{j=i+1}^{k} I\left(X_{[i]i1} > X_{[j]j1}\right),$$

where $I(.)$ is an indicator function (Vock and Balakrishnan, 2011).

The hypothesis of perfect ranking is rejected by the above test statistics for large enough values of them.

Li and Balakrishnan (2008) proposed the above test statistics in different (but equivalent) form. They then extended the above three single-cycle test statistics to the following six test statistics for the multi-cycle case.

$$N_{k,n} = \sum_{l=1}^{n} N_{kl} \qquad N^*_{k,n} = Max\left(N_{k1}, N_{k2}, N_{k3}, \cdots, N_{kn}\right)$$

$$A_{k,n} = \sum_{l=1}^{n} A_{kl} \qquad A^*_{k,n} = Max\left(A_{k1}, A_{k2}, A_{k3}, \cdots, A_{kn}\right)$$

$$S_{k,n} = \sum_{l=1}^{n} S_{kl} \qquad S^*_{k,n} = Max\left(S_{k1}, S_{k2}, S_{k3}, \cdots, S_{kn}\right)$$

where $N_{kl} = \sum_{i=1}^{k-1} \sum_{j=i+1}^{k} I\left(X_{[i]il} > X_{[j]jl}\right), S_{kl} = \sum_{i=1}^{k} \left(R_{il} - i\right)^2$, $A_{kl} = \sum_{i=1}^{k} \left|R_{il} - i\right|$ and $R_{il}$ is the rank of $X_{[i]il}$ among all measured values of the *l*th cycle and $I(.)$ is an indicator function.

Our task in this paper is to extract more information from the data in order to improve upon the tests proposed by Li and Balakrishnan (2008) in the multi-cycle case.

Suppose that

$$\begin{array}{l} X_{[1]11}, X_{[2]21}, X_{[3]31}, \cdots, X_{[k]k1} \\ X_{[1]12}, X_{[2]22}, X_{[3]32}, \cdots, X_{[k]k2} \\ X_{[1]13}, X_{[2]23}, X_{[3]33}, \cdots, X_{[k]k3} \\ \quad\vdots \qquad \vdots \qquad \vdots \qquad \vdots\ \ \vdots \\ X_{[1]1n}, X_{[2]2n}, X_{[3]3n}, \cdots, X_{[k]kn} \end{array}$$

are samples drawn by an *n*-cycle BRSS scheme. It is worth mentioning that all observations in different rows and columns are independent from each other. So, provided the assumption of perfect ranking is satisfied, it is expected that the ranks vector of observation in each row should be close to $(1,2,3,\cdots,k)$

On the other hand, since the observations in each column are independent and identically distributed, it is anticipated that the vector of $(1,2,3,\cdots,k)$ remains close to the ranks vector of the observations in the each row if we permute the observations in each column; namely, the rank vector of the vector $\left(X_{[1]1l_1}, X_{[2]2l_2}, X_{[3]3l_3},\cdots, X_{[k]kl_k}\right)$ should be close to $(1,2,3,\cdots,k)$ for all $l_1,l_2,l_3,\cdots,l_k \in \{1,2,3,\cdots,n\}$.

Based on the above discussion, we propose the following test statistics for testing perfect ranking:

$$PN_{k,n} = \sum_{h=1}^{n^k} N_{kh}^*; \qquad (1)$$

$$PA_{k,n} = \sum_{h=1}^{n^k} A_{kh}^*; \qquad (2)$$

$$PS_{k,n} = \sum_{h=1}^{n^k} S_{kh}^*; \qquad (3)$$

where $N_{kh}^*$, $A_{kh}^*$ and $S_{kh}^*$ differ from $N_{kl}, A_{kl}, S_{kl}$ in that the former set are applied on the $h$th sample out of the $n^k$ samples $\left(X_{[1]1l_1}, X_{[2]2l_2}, X_{[3]3l_3},\cdots, X_{[k]kl_k}\right)$, $l_1,l_2,l_3,\cdots,l_k \in \{1,2,3,\cdots,n\}$, instead of the $l$th cycle.

We reject the null hypothesis of perfect ranking for large enough values of the above test statistics.

**Remark 1**: It should be noted that although each observation in the $n^k$ samples $\left(X_{[1]1l_1}, X_{[2]2l_2}, X_{[3]3l_3},\cdots, X_{[k]kl_k}\right)$, $l_1,l_2,l_3,\cdots,l_k \in \{1,2,3,\cdots,n\}$, is independent from the other observations in the same sample, samples are not necessarily independent from each other anymore.

**Remark 2**: It is easy to check that the proposed test statistic $PN_{k,n}$ is equivalent to the test statistic $J_{k,n}$ proposed by Vock and Balakrishnan (2011). Therefore we use their results for this test statistic in the rest of our paper.

In addition, the test based on $PS_{k,n}$ is equivalent to the test proposed by Frey et al (2007) that is based on $W^* = \sum_{j=1}^{k}\sum_{l=1}^{n} jR_{[j]l}$, where $R_{[j]l}$ is the rank of $X_{[j]jl}$ among all $kn$ measured observations, and that rejects the hypothesis of perfect ranking for small enough values of $W^*$. To see this, we define $R_{[j]l,r}$ as the rank of $X_{[j]jl}$ in the $r$th of those permuted samples that include $X_{[j]jl}$, then $PS_{k,n}$ can be rewritten as:

$$PS_{k,n} = \sum_{r=1}^{n^{k-1}} \sum_{l=1}^{n} \sum_{j=1}^{k} \left( R_{[j]l,r} - j \right)^2 = \sum_{r=1}^{n^{k-1}} \sum_{l=1}^{n} \sum_{j=1}^{k} R_{[j]l,r}^{\;2} + \sum_{r=1}^{n^{k-1}} \sum_{l=1}^{n} \sum_{j=1}^{k} j^2 - 2\sum_{r=1}^{n^{k-1}} \sum_{l=1}^{n} \sum_{j=1}^{k} jR_{[j]l,r}$$

$$= C - 2\sum_{l=1}^{n} \sum_{j=1}^{k} j \left( \sum_{r=1}^{n^{k-1}} R_{[j]l,r} \right)$$

in which $C$ is constant and $\sum_{r=1}^{n^{k-1}} R_{[j]l,r}$ is the sum of ranks of $X_{[j]jl}$ in all $n^{k-1}$ permuted samples that include $X_{[j]jl}$. Now, it is fairly obvious that $\sum_{r=1}^{n^{k-1}} R_{[j]l,r}$ is a strictly increasing function of $R_{[j]l}$: If $R_{[j]l}$ increases by one unit, then there exists at least one permutation in which the rank of $X_{[j]jl}$ increases by one unit, so $\sum_{r=1}^{n^{k-1}} R_{[j]l,r}$ also will be increased. Therefore tests based on $PS_{k,n}$ and $W^*$ are equivalent.

**Remark 3:** It is worth mentioning that Vock and Balakrishnan (2012) extended $S_k$ to the multi-cycle case in a different way which turns out to be equivalent to the test statistic proposed by Frey et al (2007), and therefore it is also equivalent to $PS_{k,n}$.

In order to test the hypothesis of perfect ranking, we need to obtain the null distribution of the proposed test statistic. Although it is possible to obtain the null distribution of the proposed test statistic in (2) in the light of Theorem 1 in Frey (2007), it is not practical for large values of *k*, *n*, because the number of different probabilities to be calculated grows rapidly in *k* and *n* (see Vock and Balakrishnan, 2011, for more details about the derivation of such exact distributions). Thus we have to obtain some of the critical values of the test statistics by using Monte Carlo simulation.

Tables 1 shows critical values of $PA_{k,n}$ as well as the corresponding significance levels close to 0.05 and 0.10 for $k = 2,3,4,5$ and $n = 2,3,4,5$. Those entries marked by asterisks were obtained by using Monte Carlo simulation with 100,000 repetitions, while the others are exact.

**Table 1** Critical values (CV) of the test based on $PA_{k,n}$ for nominal levels close to 0.05 and 0.10, and the corresponding levels, $P_0\left(PA_{k,n} \geq CV\right)$.

| n | k=2 | | k=3 | | k=4 | | k=5 | |
|---|---|---|---|---|---|---|---|---|
| | CV | Level | CV | Level | CV | Level | CV | Level |
| 2 | 6 | .04127 | 20 | .03995 | 54 | .04880 | 148 | .04201 |
| | 4 | .19683 | 18 | .05101 | 52 | .06134 | 144 | .06366 |
| | | | 16 | .12158 | 50 | .06210 | 132 | .08340 |
| | | | | | 48 | .11186 | 128 | .12514 |
| 3 | 12 | .02587 | 54 | .04707 | 238 | .04939 | 986 | .04955* |
| | 10 | .05527 | 52 | .05473 | 236 | .05019 | 984 | .05081* |
| | 8 | .12016 | 46 | .09855 | 212 | .09783 | 892 | .09865* |
| | | | 44 | .10698 | 210 | .10929 | 888 | .10081* |
| 4 | 16 | .04579 | 114 | .04489 | 698 | .04985* | 3906 | .04838* |
| | 14 | .07721 | 112 | .05519 | 696 | .05216* | 3904 | .05015* |
| | 12 | .13444 | 98 | .09901 | 626 | .09877* | 3576 | .09904* |
| | | | 96 | .12028 | 624 | .10503* | 3568 | .10024* |
| 5 | 22 | .04902 | 208 | .04733 | 1618 | .04999* | 11394 | .04994* |
| | 20 | .07515 | 206 | .05038 | 1616 | .05022* | 11390 | .05023* |
| | 18 | .11089 | 182 | .09848 | 1468 | .09977* | 10504 | .09980* |
| | | | 180 | .10898 | 1466 | .10003* | 10500 | .10117 |

*The levels with asterisks are obtained by using Monte Carlo simulation with 100,000 repetitions, while the others are exact.

Comparing the critical values and exact levels of $PA_{k,n}$ (Table 1) with those of $J_{k,n}\left(PN_{k,n}\right)$ (Table 1 of Vock and Balakrishnan, 2011) and $W^*\left(PS_{k,n}\right)$ (Table 4 of Frey et al, 2007), it is interesting to see that in many cases, $PA_{k,n}$ offers exact levels much closer to 0.05 (or any other desired level) than those offered by $J_{k,n}\left(PN_{k,n}\right)$ and $W^*\left(PS_{k,n}\right)$ since the null distribution of $PA_{k,n}$ typically has smaller point probabilities. While we use randomized tests in the simulation study in Section 3, this fact can lead to a relevant advantage in power in the practical application of non-randomized tests, mainly in cases with small values of *k* and *n*.

## 3. Power Comparisons

In this section, we compare our new proposed test based on $PA_{k,n}$ with its leading competitors, namely:

- The test based on $N_{k,n}$ from Li and Balakrishnan (2008)
- The test based on $S_{k,n}$ from Li and Balakrishnan (2008)
- The test based on $A_{k,n}$ from Li and Balakrishnan (2008).

- The test based on $W^*$ (equivalent to $PS_{k,n}$) from Frey et al (2007).
- The test based on $J_{k,n}$ (equivalent to $PN_{k,n}$) from Vock and Balakrishnan (2011).

It should be mentioned that since Li and Balakrishnan's (2008) simulation results confirm that $N^*_{k,n}, A^*_{k,n}, S^*_{k,n}$ are uniformly less powerful than $N_{k,n}, A_{k,n}, S_{k,n}$, we leave them out of the comparison and we do not use this technique in proposing new tests as well.
For the power comparison, we use the following four different scenarios of imperfect ranking which were proposed by Frey et al (2007) and Vock and Balakrishnan (2011).

- *Concomitant*. Suppose that $(X,Y)$ have a bivariate normal distribution with correlation coefficient $\lambda$. We rank the variable of interest $X$ according to the true order of the concomitant variable ($Y$).
- *Fraction of random rankings*: With probability $\lambda$, we replace $X_{[i]il}$ by a new observation from the original distribution, otherwise we obtain it by perfect ranking, so the distribution function of $X_{[i]il}$ will be $F_{[i]} = (1-\lambda)F_{(i)} + \lambda F$, for $i = 1,\cdots,k$, some $\lambda \in [0,1]$.
- *Fraction of inverse rankings:* With probability $\lambda$, we replace $X_{[i]il}$ by $X_{(k-i+1)il}$, otherwise we obtain it by perfect ranking, so the distribution function of $X_{[i]il}$ will be $F_{[i]} = (1-\lambda)F_{(i)} + \lambda F_{(k-i+1)}$, for $i = 1,\cdots,k$, some $\lambda \in [0,1]$.
- *Fraction of neighbors:* With probability $\frac{\lambda}{2}$, we substitute $X_{[i]il}$ by $X_{(i-1)il}$, and with the same probability we substitute it by $X_{(i+1)il}$ (using $X_{(0)il} = X_{(1)il}$ and $X_{(k+1)il} = X_{(k)il}$), otherwise we obtain $X_{[i]il}$ by perfect ranking. Thus the distribution function of $X_{[i]il}$ will be $F_{[i]} = \frac{\lambda}{2}F_{(i-1)} + (1-\lambda)F_{(i)} + \frac{\lambda}{2}F_{(i+1)}$, for $i = 1,\cdots,k$, some $\lambda \in [0,1]$ and $F_{(0)} = F_{(1)}, F_{(k+1)} = F_{(k)}$.

It should be mentioned that we directly report the powers of the all competing tests from Vock and Balakrishnan (2011) (Tables 3-6), whereas new simulations are done for the powers of the test based on $PA_{k,n}$. The powers of these tests are estimated based on 100,000 repetitions of Monte Carlo simulation under the above scenarios of imperfect ranking for different values of $\lambda$, ($n$,$k$) at a significance level of $\alpha = 0.05$ using randomized tests. The results are tabulated in Tables 2-5.

**Remark 4:** It is easy to check that all tests based on $W^*(PS_{k,n}), PA_{k,n}, J_{k,n}(PN_{k,n})$ are equivalent for k=2 and all values of n).

***Table 2**. Power estimates of the tests based on* $N_{k,n}$, $S_{k,n}$, $A_{k,n}$, $J_{k,n}\left(PN_{k,n}\right)$, $W^*\left(PS_{k,n}\right)$ and $PA_{k,n}$ *under the concomitant model with correlation* $\lambda$ *at significance level* $\alpha = 0.05$

| $k$ | $n$ | Test Statistics | $\lambda$ | | | | | | | | | | |
|---|---|---|---|---|---|---|---|---|---|---|---|---|---|
| | | | .5 | .55 | .60 | .65 | .70 | .75 | .80 | .85 | .90 | .95 | 1 |
| 2 | 5 | $N_{k,n}$ | .2468 | .2244 | .2010 | .1771 | .1530 | .1327 | .1132 | .0950 | .0791 | .0629 | .0493 |
| 2 | 5 | $S_{k,n}$ | .2468 | .2244 | .2010 | .1771 | .1530 | .1327 | .1132 | .0950 | .0791 | .0629 | .0493 |
| 2 | 5 | $A_{k,n}$ | .2468 | .2244 | .2010 | .1771 | .1530 | .1327 | .1132 | .0950 | .0791 | .0629 | .0493 |
| 2 | 5 | $J_{k,n}\left(PN_{k,n}\right)$ | .3180 | .2838 | .2507 | .2215 | .1890 | .1631 | .1353 | .1092 | .0878 | .0663 | .0506 |
| 2 | 5 | $W^*\left(PS_{k,n}\right)$ | .3181 | .2839 | .2508 | .2216 | .1890 | .1632 | .1353 | .1093 | .0878 | .0663 | .0506 |
| 2 | 5 | $PA_{k,n}$ | .3180 | .2838 | .2507 | .2215 | .1890 | .1631 | .1353 | .1092 | .0878 | .0663 | .0506 |
| | | | | | | | | | | | | | |
| 5 | 2 | $N_{k,n}$ | .6820 | .6345 | .5799 | .5202 | .4560 | .3878 | .3140 | .2386 | .1694 | .1038 | .0503 |
| 5 | 2 | $S_{k,n}$ | .6982 | .6498 | .5942 | .5339 | .4672 | .3995 | .3259 | .2521 | .1742 | .1074 | .0511 |
| 5 | 2 | $A_{k,n}$ | .6689 | .6213 | .5693 | .5107 | .4489 | .3826 | .3096 | .2371 | .1677 | .1037 | .0499 |
| 5 | 2 | $J_{k,n}\left(PN_{k,n}\right)$ | .7366 | .6921 | .6367 | .5765 | .5086 | .4351 | .3529 | .2669 | .1877 | .1113 | .0503 |
| 5 | 2 | $W^*\left(PS_{k,n}\right)$ | .7480 | .6990 | .6488 | .5857 | .5188 | .4402 | .3624 | .2772 | .1924 | .1133 | .0508 |
| 5 | 2 | $PA_{k,n}$ | .7429 | .6965 | .6417 | .5795 | .5091 | .4332 | .3560 | .2747 | .1894 | .1115 | .0514 |
| | | | | | | | | | | | | | |
| 4 | 5 | $N_{k,n}$ | .8047 | .7555 | .6930 | .6262 | .5457 | .4622 | .3715 | .2754 | .1856 | .1075 | .0506 |
| 4 | 5 | $S_{k,n}$ | .8083 | .7588 | .6965 | .6307 | .5516 | .4641 | .3728 | .2756 | .1878 | .1085 | .0488 |
| 4 | 5 | $A_{k,n}$ | .7827 | .7293 | .6683 | .6026 | .5244 | .4413 | .3526 | .2632 | .1791 | .1053 | .0499 |
| 4 | 5 | $J_{k,n}\left(PN_{k,n}\right)$ | .8876 | .8454 | .7961 | .7299 | .6510 | .5577 | .4517 | .3403 | .2245 | .1252 | .0508 |
| 4 | 5 | $W^*\left(PS_{k,n}\right)$ | .8846 | .8431 | .7913 | .7281 | .6496 | .5570 | .4510 | .3367 | .2234 | .1235 | .0511 |
| 4 | 5 | $PA_{k,n}$ | .8865 | .8440 | .7938 | .7276 | .6499 | .5513 | .4476 | .3329 | .2201 | .1225 | .0487 |
| | | | | | | | | | | | | | |
| 5 | 4 | $N_{k,n}$ | .8877 | .8504 | .7988 | .7392 | .6644 | .5770 | .4688 | .3533 | .2381 | .1282 | .0503 |
| 5 | 4 | $S_{k,n}$ | .8994 | .8621 | .8158 | .7544 | .6801 | .5935 | .4822 | .3689 | .2440 | .1355 | .0507 |
| 5 | 4 | $A_{k,n}$ | .8810 | .8433 | .7918 | .7316 | .6525 | .5636 | .4592 | .3471 | .2307 | .1287 | .0512 |
| 5 | 4 | $J_{k,n}\left(PN_{k,n}\right)$ | .9439 | .9175 | .8805 | .8338 | .7639 | .6752 | .5629 | .4316 | .2856 | .1514 | .0510 |
| 5 | 4 | $W^*\left(PS_{k,n}\right)$ | .9448 | .9175 | .8792 | .8326 | .7649 | .6754 | .5646 | .4324 | .2883 | .1513 | .0494 |
| 5 | 4 | $PA_{k,n}$ | .9431 | .9174 | .8807 | .8315 | .7635 | .6739 | .5632 | .4277 | .2855 | .1478 | .0496 |

***Table 3**. Power estimates of the tests based on* $N_{k,n}$, $S_{k,n}$, $A_{k,n}$, $J_{k,n}\left(PN_{k,n}\right)$, $W^*\left(PS_{k,n}\right)$ and $PA_{k,n}$ *under the fraction of random rankings model with fraction* $\lambda$ *at significance level* $\alpha = 0.05$

| $k$ | $n$ | Test Statistics | $\lambda$ | | | | | | | | | | |
|---|---|---|---|---|---|---|---|---|---|---|---|---|---|
| | | | 0 | .05 | 0.1 | 0.15 | 0.2 | 0.25 | 0.3 | 0.35 | 0.4 | 0.45 | 0.5 |
| 2 | 5 | $N_{k,n}$ | .0501 | .0618 | .0756 | .0908 | .1092 | .1268 | .1474 | .1681 | .1925 | .2170 | .2403 |
| 2 | 5 | $S_{k,n}$ | .0501 | .0618 | .0756 | .0908 | .1092 | .1268 | .1474 | .1681 | .1925 | .2170 | .2403 |
| 2 | 5 | $A_{k,n}$ | .0501 | .0618 | .0756 | .0908 | .1092 | .1268 | .1474 | .1681 | .1925 | .2170 | .2403 |
| 2 | 5 | $J_{k,n}\left(PN_{k,n}\right)$ | .0502 | .0651 | .0847 | .1077 | .1296 | .1541 | .1789 | .2102 | .2404 | .2716 | .3067 |
| 2 | 5 | $PA_{k,n}$ | .0502 | .0651 | .0847 | .1077 | .1296 | .1541 | .1789 | .2102 | .2404 | .2716 | .3067 |
| 2 | 5 | $W^*\left(PS_{k,n}\right)$ | .0503 | .0652 | .0847 | .1078 | .1296 | .1541 | .1790 | .2102 | .2404 | .2717 | .3068 |
| | | | | | | | | | | | | | |
| 5 | 2 | $N_{k,n}$ | .0484 | .1015 | .1595 | .2199 | .2815 | .3456 | .4061 | .4653 | .5242 | .5783 | .6304 |
| 5 | 2 | $S_{k,n}$ | .0500 | .1143 | .1795 | .2447 | .3120 | .3749 | .4402 | .4998 | .5545 | .6080 | .6579 |
| 5 | 2 | $A_{k,n}$ | .0506 | .0987 | .1499 | .2062 | .2660 | .3245 | .3851 | .4424 | .4995 | .5530 | .6046 |
| 5 | 2 | $J_{k,n}\left(PN_{k,n}\right)$ | .0500 | .1117 | .1758 | .2433 | .3123 | .3786 | .4481 | .5120 | .5719 | .6276 | .6775 |
| 5 | 2 | $W^*\left(PS_{k,n}\right)$ | .0518 | .1233 | .1988 | .2708 | .3443 | .4112 | .4801 | .5408 | .6017 | .6550 | .7024 |
| 5 | 2 | $PA_{k,n}$ | .0496 | .1108 | .1713 | .2383 | .3062 | .3719 | .4410 | .5067 | .5651 | .6194 | .6750 |
| | | | | | | | | | | | | | |
| 4 | 5 | $N_{k,n}$ | .0500 | .1043 | .1701 | .2469 | .3260 | .4071 | .4854 | .5629 | .6332 | .7001 | .7588 |
| 4 | 5 | $S_{k,n}$ | .0502 | .1137 | .1891 | .2712 | .3529 | .4364 | .5179 | .5923 | .6594 | .7231 | .7768 |
| 4 | 5 | $A_{k,n}$ | .0504 | .0985 | .1584 | .2276 | .2960 | .3764 | .4536 | .5264 | .5976 | .6637 | .7200 |
| 4 | 5 | $J_{k,n}\left(PN_{k,n}\right)$ | .0504 | .1204 | .2047 | .2959 | .3934 | .4912 | .5767 | .6572 | .7291 | .7873 | .8376 |
| 4 | 5 | $W^*\left(PS_{k,n}\right)$ | .0488 | .1312 | .2260 | .3257 | .4238 | .5173 | .6041 | .6817 | .7468 | .8042 | .8505 |
| 4 | 5 | $PA_{k,n}$ | .0501 | .1141 | .1931 | .2808 | .3760 | .4712 | .5585 | .6380 | .7124 | .7750 | .8255 |
| | | | | | | | | | | | | | |
| 5 | 4 | $N_{k,n}$ | .0492 | .1196 | .2057 | .3007 | .3969 | .4900 | .5775 | .6535 | .7248 | .7845 | .8362 |
| 5 | 4 | $S_{k,n}$ | .0488 | .1418 | .2462 | .3485 | .4492 | .5416 | .6294 | .7022 | .7610 | .8202 | .8618 |
| 5 | 4 | $A_{k,n}$ | .0488 | .1141 | .1908 | .2814 | .3732 | .4650 | .5513 | .6300 | .6998 | .7650 | .8157 |
| 5 | 4 | $J_{k,n}\left(PN_{k,n}\right)$ | .0483 | .1412 | .2480 | .3596 | .4700 | .5716 | .6635 | .7425 | .8061 | .8567 | .8966 |
| 5 | 4 | $W^*\left(PS_{k,n}\right)$ | .0488 | .1644 | .2814 | .3987 | .5144 | .6116 | .6962 | .7706 | .8260 | .8725 | .9090 |
| 5 | 4 | $PA_{k,n}$ | .0498 | .1341 | .2351 | .3435 | .4517 | .5561 | .6445 | .7250 | .7937 | .8472 | .8887 |

***Table 4**. Power estimates of the tests based on* $N_{k,n}$, $S_{k,n}$, $A_{k,n}$, $J_{k,n}\left(PN_{k,n}\right)$, $W^{*}\left(PS_{k,n}\right)$ and $PA_{k,n}$ *under the fraction of inverse rankings model with fraction* $\lambda$ *at significance level* $\alpha = 0.05$

| $k$ | $n$ | Test Statistics | $\lambda$ | | | | | | | | | | |
|---|---|---|---|---|---|---|---|---|---|---|---|---|---|
| | | | 0 | .05 | 0.1 | 0.15 | 0.2 | 0.25 | 0.3 | 0.35 | 0.4 | 0.45 | 0.5 |
| 2 | 5 | $N_{k,n}$ | .0501 | .0760 | .1082 | .1437 | .1892 | .2405 | .2930 | .3475 | .4051 | .4650 | .5301 |
| 2 | 5 | $S_{k,n}$ | .0501 | .0760 | .1082 | .1437 | .1892 | .2405 | .2930 | .3475 | .4051 | .4650 | .5301 |
| 2 | 5 | $A_{k,n}$ | .0501 | .0760 | .1082 | .1437 | .1892 | .2405 | .2930 | .3475 | .4051 | .4650 | .5301 |
| 2 | 5 | $J_{k,n}\left(PN_{k,n}\right)$ | .0502 | .0859 | .1237 | .1819 | .2421 | .3061 | .3776 | .4535 | .5191 | .5905 | .6590 |
| 2 | 5 | $W^{*}\left(PS_{k,n}\right)$ | .0503 | .0859 | .1237 | .1820 | .2422 | .3061 | .3777 | .4535 | .5192 | .5905 | .6590 |
| 2 | 5 | $PA_{k,n}$ | .0502 | .0859 | .1237 | .1819 | .2421 | .3061 | .3776 | .4535 | .5191 | .5905 | .6590 |
| | | | | | | | | | | | | | |
| 5 | 2 | $N_{k,n}$ | .0484 | .1589 | .2745 | .3873 | .4888 | .5856 | .6740 | .7482 | .8114 | .8620 | .9029 |
| 5 | 2 | $S_{k,n}$ | .0500 | .1890 | .3183 | .4402 | .5452 | .6378 | .7191 | .7876 | .8439 | .8850 | .9205 |
| 5 | 2 | $A_{k,n}$ | .0506 | .1499 | .2524 | .3544 | .4520 | .5477 | .6317 | .7052 | .7711 | .8264 | .8729 |
| 5 | 2 | $J_{k,n}\left(PN_{k,n}\right)$ | .0500 | .1757 | .3014 | .4224 | .5307 | .6275 | .7114 | .7842 | .8419 | .8878 | .9246 |
| 5 | 2 | $W^{*}\left(PS_{k,n}\right)$ | .0518 | .2045 | .3452 | .4722 | .5799 | .6770 | .7518 | .8172 | .8678 | .9075 | .9380 |
| 5 | 2 | $PA_{k,n}$ | .0506 | .1710 | .2925 | .4071 | .5144 | .6089 | .6958 | .7673 | .8273 | .8748 | .9134 |
| | | | | | | | | | | | | | |
| 4 | 5 | $N_{k,n}$ | .0500 | .1685 | .3128 | .4615 | .5961 | .7150 | .8078 | .8761 | .9241 | .9556 | .9776 |
| 4 | 5 | $S_{k,n}$ | .0502 | .2007 | .3610 | .5178 | .6499 | .7566 | .8403 | .8994 | .9397 | .9659 | .9816 |
| 4 | 5 | $A_{k,n}$ | .0504 | .1516 | .2786 | .4129 | .5423 | .6616 | .7573 | .8348 | .8921 | .9358 | .9629 |
| 4 | 5 | $J_{k,n}\left(PN_{k,n}\right)$ | .0504 | .2020 | .3754 | .5374 | .6761 | .7866 | .8675 | .9226 | .9572 | .9768 | .9893 |
| 4 | 5 | $W^{*}\left(PS_{k,n}\right)$ | .0488 | .2377 | .4283 | .5933 | .7246 | .8221 | .8913 | .9380 | .9666 | .9828 | .9914 |
| 4 | 5 | $PA_{k,n}$ | .0503 | .1887 | .3511 | .5142 | .6532 | .7667 | .8494 | .9091 | .9477 | .9719 | .9860 |
| | | | | | | | | | | | | | |
| 5 | 4 | $N_{k,n}$ | .0492 | .2006 | .3698 | .5329 | .6747 | .7828 | .8600 | .9180 | .9547 | .9750 | .9878 |
| 5 | 4 | $S_{k,n}$ | .0488 | .2565 | .4526 | .6173 | .7398 | .8356 | .8994 | .9428 | .9686 | .9848 | .9926 |
| 5 | 4 | $A_{k,n}$ | .0488 | .1834 | .3381 | .4905 | .6281 | .7426 | .8280 | .8924 | .9361 | .9640 | .9813 |
| 5 | 4 | $J_{k,n}\left(PN_{k,n}\right)$ | .0483 | .2398 | .4356 | .6079 | .7410 | .8387 | .9052 | .9481 | .9731 | .9880 | .9944 |
| 5 | 4 | $W^{*}\left(PS_{k,n}\right)$ | .0495 | .2941 | .5075 | .6740 | .7920 | .8761 | .9295 | .9625 | .9811 | .9910 | .9958 |
| 5 | 4 | $PA_{k,n}$ | .0499 | .2221 | .4070 | .5776 | .7141 | .8182 | .8912 | .9381 | .9675 | .9839 | .9927 |

***Table 5**. Power estimates of the tests based on* $N_{k,n}$, $S_{k,n}$, $A_{k,n}$, $J_{k,n}\left(PN_{k,n}\right)$, $W^*\left(PS_{k,n}\right)$ and $PA_{k,n}$ *under the fraction of neighbors model with fraction* $\lambda$ *at significance level* $\alpha = 0.05$

| $k$ | $n$ | Test Statistics | $\lambda$ | | | | | | | | | | |
|---|---|---|---|---|---|---|---|---|---|---|---|---|---|
| | | | 0 | 0.1 | 0.2 | 0.3 | 0.4 | 0.5 | 0.6 | 0.7 | 0.8 | 0.9 | 1 |
| 2 | 5 | $N_{k,n}$ | .0501 | .0758 | .1079 | .1463 | .1917 | .2387 | .2910 | .3490 | .4064 | .4698 | .5269 |
| 2 | 5 | $S_{k,n}$ | .0501 | .0758 | .1079 | .1463 | .1917 | .2387 | .2910 | .3490 | .4064 | .4698 | .5269 |
| 2 | 5 | $A_{k,n}$ | .0501 | .0758 | .1079 | .1463 | .1917 | .2387 | .2910 | .3490 | .4064 | .4698 | .5269 |
| 2 | 5 | $J_{k,n}\left(PN_{k,n}\right)$ | .0502 | .0851 | .1285 | .1821 | .2413 | .3054 | .3754 | .4466 | .5179 | .5894 | .6562 |
| 2 | 5 | $W^*\left(PS_{k,n}\right)$ | .0503 | .0852 | .1285 | .1822 | .2413 | .3055 | .3755 | .4467 | .5179 | .5895 | .6562 |
| 2 | 5 | $PA_{k,n}$ | .0502 | .0851 | .1285 | .1821 | .2413 | .3054 | .3754 | .4466 | .5179 | .5894 | .6562 |
| | | | | | | | | | | | | | |
| 5 | 2 | $N_{k,n}$ | .0484 | .0673 | .0861 | .1081 | .1346 | .1562 | .1873 | .2140 | .2464 | .2770 | .3086 |
| 5 | 2 | $S_{k,n}$ | .0500 | .0660 | .0868 | .1080 | .1309 | .1551 | .1792 | .2098 | .2334 | .2666 | .2930 |
| 5 | 2 | $A_{k,n}$ | .0506 | .0681 | .0881 | .1117 | .1377 | .1647 | .1940 | .2248 | .2543 | .2919 | .3299 |
| 5 | 2 | $J_{k,n}\left(PN_{k,n}\right)$ | .0500 | .0694 | .0930 | .1189 | .1437 | .1750 | .2046 | .2418 | .2777 | .3148 | .3505 |
| 5 | 2 | $W^*\left(PS_{k,n}\right)$ | .0518 | .0703 | .0916 | .1168 | .1420 | .1698 | .2001 | .2317 | .2616 | .2973 | .3291 |
| 5 | 2 | $PA_{k,n}$ | .0506 | .0726 | .0945 | .1223 | .1543 | .1873 | .2226 | .2605 | .2974 | .3391 | .3827 |
| | | | | | | | | | | | | | |
| 4 | 5 | $N_{k,n}$ | .0500 | .0798 | .1178 | .1592 | .2079 | .2634 | .3205 | .3818 | .4444 | .5066 | .5674 |
| 4 | 5 | $S_{k,n}$ | .0502 | .0765 | .1133 | .1535 | .1998 | .2477 | .3010 | .3592 | .4154 | .4728 | .5274 |
| 4 | 5 | $A_{k,n}$ | .0504 | .0771 | .1136 | .1580 | .2064 | .2615 | .3224 | .3836 | .4482 | .5120 | .5748 |
| 4 | 5 | $J_{k,n}\left(PN_{k,n}\right)$ | .0504 | .0866 | .1341 | .1892 | .2538 | .3229 | .3980 | .4762 | .5509 | .6254 | .6939 |
| 4 | 5 | $W^*\left(PS_{k,n}\right)$ | .0488 | .0842 | .1292 | .1816 | .2423 | .3062 | .3730 | .4428 | .5106 | .5796 | .6430 |
| 4 | 5 | $PA_{k,n}$ | .0501 | .0865 | .1320 | .1923 | .2597 | .3349 | .4123 | .4944 | .5731 | .6493 | .7156 |
| | | | | | | | | | | | | | |
| 5 | 4 | $N_{k,n}$ | .0492 | .0742 | .1045 | .1407 | .1805 | .2258 | .2718 | .3214 | .3775 | .4298 | .4812 |
| 5 | 4 | $S_{k,n}$ | .0488 | .0744 | .1029 | .1364 | .1750 | .2172 | .2608 | .3058 | .3555 | .4065 | .4530 |
| 5 | 4 | $A_{k,n}$ | .0488 | .0750 | .1060 | .1434 | .1867 | .2333 | .2829 | .3386 | .3955 | .4510 | .5105 |
| 5 | 4 | $J_{k,n}\left(PN_{k,n}\right)$ | .0483 | .0784 | .1190 | .1635 | .2159 | .2750 | .3349 | .4004 | .4659 | .5369 | .5949 |
| 5 | 4 | $W^*\left(PS_{k,n}\right)$ | .0495 | .0804 | .1136 | .1569 | .2048 | .2557 | .3103 | .3695 | .4285 | .4858 | .5485 |
| 5 | 4 | $PA_{k,n}$ | .0503 | .0821 | .1205 | .1694 | .2250 | .2901 | .3578 | .4265 | .4973 | .5693 | .6316 |

The simulation results are reasonably good in favor of the proposed tests. The proposed tests are uniformly better than their counterparts proposed by Li and Balakrishnan (2008) in all different scenarios and for all considered values of $\lambda$, ($n$,$k$). Table 3 indicates that under the concomitant model, the proposed tests are quite competitive with each other and there are no considerable differences between their powers. Under the fraction of random and inverse rankings models (Tables 4-5), the results are more in favor of the tests based on $W^*\left(PS_{k,n}\right)$. When the fraction of neighbors is the model of imperfect ranking (Table 6), the test based on $PA_{k,n}$ is uniformly better than the other tests for all considered values of $\lambda$, ($n$,$k$), where its superiority is not negligible.

**Remark 5**: Although the test statistic $PA_{k,n}$ can be easily calculated for small and moderate sample sizes by looking at all possible permutations, this would become too time-consuming for large values of n, k. An anonymous referee suggested the following technique in order to reduce the calculation burden of computing $PA_{k,n}$: Let $p_i$ be the proportion of values with in-set rank i that are smaller than $X_{[j]jl}$. It then follows that when we select at random a permutation that includes $X_{[j]jl}$, the rank for $X_{[j]jl}$ (here denoted by $R^*_{j,l}$) is distributed like $1+B_1+\cdots+B_{j-1}+B_{j+1}+\cdots+B_k$, where $B_i$ is Bernoulli with success parameter $p_i$ and the k-1 random variables are mutually independent, such that the distribution of the sum can easily be obtained by a convolution of the Bernoulli distributions. The test statistic can then be calculated by: $PA_{k,n}=n^{k-1}\left(\sum_{l=1}^{n}\sum_{i=1}^{k}E\left|R^*_{i,l}-i\right|\right)$.

## 4. Conclusion

In this paper, we proposed three test statistics for testing the hypothesis of perfect ranking in balanced ranked set sampling. Our proposed tests are derived based on all permutations of independent and identically distributed observations in the BRSS sample. We also mentioned that two of the proposed tests are equivalent to the tests proposed by Vock and Balakrishnan (2011) and Frey et al (2007). The following observations are made based on our comparisons:

I. Each proposed test is uniformly better than its counterpart proposed by Li and Balakrishnan (2008) in all different scenarios of imperfect rankings and for all considered values of $\lambda$, ($n$,$k$).
II. All proposed tests are recommendable to be used in practice if potential ranking errors can be assumed to follow the concomitant model.
III. If severe ranking errors are plausible (such as under the fraction of random and inverse rankings models), the test based on $W^*\left(PS_{k,n}\right)$ should be used in practice.
IV. When small ranking errors are most plausible (such as in the fraction of neighbors model), usage of the test based on $PA_{k,n}$ should be in priority.

We assume that in practical applications of tests for perfect ranking, a researcher will often be aware of some possible sources of ranking errors that are most plausible, and he will therefore be able to make a suitable choice. E.g., if a ranking is actually based on a concomitant variable that is measured exactly, then the concomitant model may be most plausible. It is worth mentioning that although the fraction of neighbors model of imperfect rankings leads to mild ranking errors, it seems more likely to occur in practice than models based on random and inverse ranking.

Our simulations were based on randomized tests. Since the non-randomized tests based $PA_{k,n}$ can usually be performed at a size closer to the desired level than the other tests investigated, the loss of power due to the use of non-randomized tests will be smaller in this test than in other tests. For small values of *k* and *n*, the test based on $PA_{k,n}$ could therefore be attractive in practice even if the power according to the simulations given here is not optimal among the randomized tests.

**Acknowledgements:** The authors are very thankful to an anonymous referee for comments which led to a considerable improvement to an earlier version of this paper.